# Polytypism, polymorphism and superconductivity in TaSe$_{2-x}$Te$_x$

Huixia Luo[1*], Weiwei Xie[1], Jing Tao[2], Hiroyuki Inoue[3], András Gyenis[3], Jason W. Krizan[1], Ali Yazdani[3], Yimei Zhu[2], and R. J. Cava[1*]

[1]*Department of Chemistry, Princeton University, Princeton, New Jersey 08544, USA.*

[2]*Department of Condensed Matter Physics and Materials Science, Brookhaven National Laboratory, Upton, New York 11973, USA.*

[3]*Joseph Henry Laboratories and Department of Physics, Princeton University, Princeton, New Jersey 08544, USA.*

\* rcava@princeton.edu;huixial@princeton.edu

Polymorphism in materials often leads to significantly different physical properties - the rutile and anatase polymorphs of $TiO_2$ are a prime example. Polytypism is a special type of polymorphism, occurring in layered materials when the geometry of a repeating structural layer is maintained but the layer stacking sequence of the overall crystal structure can be varied; SiC is an example of a material with many polytypes. Although polymorphs can have radically different physical properties, it is much rarer for polytypism to impact physical properties in a dramatic fashion. Here we study the effects of polytypism and polymorphism on the superconductivity of $TaSe_2$, one of the archetypal members of the large family of layered dichalcogenides. We show that it is possible to access 2 stable polytypes and 2 stable polymorphs in the $TaSe_{2-x}Te_x$ solid solution, and find that the 3R polytype shows a superconducting transition temperature that is nearly 17 times higher than that of the much more commonly found 2H polytype. The reason for this dramatic change is not apparent, but we propose that it arises either from a remarkable dependence of $T_c$ on subtle differences in the characteristics of the single layers present, or from a surprising effect of the layer stacking sequence on electronic properties that instead are expected to be dominated by the properties of a single layer in materials of this kind.

The MX$_2$ Layered transition-metal dichalcogenides (TMDCs, M = Mo, W, V, Nb, Ta, Ti, Zr, Hf, and Re, and X = Se, S or Te), have long been of interest due to the rich electronic properties that emerge due to their low dimensionality (1-9). Structurally, these compounds can be regarded as having strongly bonded (2D) X–M–X layers, with M in either trigonal prismatic or octahedral coordination with X, and weak inter-layer X-X bonding of the van der Waals type. Many of these materials manifest charge density waves and the competition between CDWs and superconductivity, e.g. (5-9). Among the TMDCs, the 2H polytype of tantalum diselenide (2H-TaSe$_2$) is considered one of the foundational materials (8-18), showing a transition from a metallic phase to an incommensurate charge-density-wave (ICDW) phase at 123 K, followed by a "lock-in" transition to a commensurate charge-density-wave (CCDW) phase at 90 K. It finally becomes a superconductor with a rather low T$_c$ of 0.15 K. Although detailed studies have been performed on the physics of CDWs and superconductivity in 2H-TaSe$_2$ (16-18), a comparative study of the superconductivity of the polytypes and polymorphs of TaSe$_2$ from the chemical perspective has not been done.

TaSe$_2$ is highly polymorphic, possibly the most polymorphic of the TMDCs (19). In some of its forms, notably the 2H and 3R polytypes (Figure 1*a*), Ta is found in trigonal prismatic coordination in Se-Ta-Se layers that are stacked along the *c* axis of the hexagonal (or rhombohedral) cell. The 2H and 3R polytypes differ only in their stacking periodicity – the structure repeats after 2 layers in the 2H form and 3 layers in the 3R form (20-22). The 3R form can be synthesized, but it is not the stable variant (the 2H form is) and so has been the subject of little study. In one of the other polymorphs, the 1T type, Ta is found in octahedral coordination in the Se-Ta-Se layers, and the layer stacking along the *c* axis of the trigonal cell such that the structure repeats after only one layer (23) (Figure 1*a*). Again, the 1T form has not been the subject of much study. Here we show that the 3R and 1T polymorphs are both quite stable in the TaSe$_{2-x}$Te$_x$ system, and that they are both superconducting. For pure TaTe$_2$, the monoclinic structure is 1T-based (Figure 1*a*), but is distorted such that there are two non-equivalent Ta and three non-equivalent Te positions in the unit cell (24); we find TaSe$_{2-x}$Te$_x$ in this polymorph to be non-superconducting down to 0.4 K.

We report the structures and superconducting properties of TaSe$_{2-x}$Te$_x$ for $0 \leq x \leq 2$. The 2H (H: hexagonal), 3R (R: rhombohedral), 1T (T: trigonal) and monoclinic distorted 1T-structure forms were successfully synthesized. Only a small amount of Te doping ($x = 0.02$) changes 2H-TaSe$_2$ into the 3R-polytype. Within the 3R polytype, TaSe$_{2-x}$Te$_x$ shows the coexistence of a CDW and superconductivity above 0.4 K for $0.1 \leq x \leq 0.35$. For x = 0.35, 3R-TaSe$_{1.65}$Te$_{0.35}$ shows the highest T$_c$ in the system, 2.4 K, which is 17 times higher than that of 2H-TaSe$_2$. For $0.8 \leq x \leq 1.3$, 1T-type TaSe$_{2-x}$Te$_x$ emerges and shows a T$_c$ of 0.5 - 0.7 K. At higher Te substitutions ($1.8 \leq x \leq 2$), TaSe$_{2-x}$Te$_x$ changes again, into the monoclinic polymorph, and shows normal metallic behavior to 0.4 K. We argue that the isovalent Te/Se substitution acts to tune the anisotropy of the layers, inducing the 3R to 1T transition, consistent with what has been proposed previously (25).The driving force for the 2H to 3R transition currently remains obscure.

**Results and Discussion**

The polycrystalline samples of TaSe$_{2-x}$Te$_x$ were prepared as described in the experimental section. In the composition range of $0.02 \leq x \leq 0.35$, the samples have the non-centrosymmetric rhombohedral 3R structure (*R3m, space group #160*), evidenced by their powder X-ray diffraction (PXRD) patterns. The 3R-stucture of the materials in this composition range is also confirmed by single crystal X-ray diffraction and electron diffraction. The detailed crystallographic data determined from the quantitative structure refinements of a single crystal of the 3R phase are summarized in Tables 1 and 2. In the refined crystal structure of 3R TaSe$_{1.7}$Te$_{0.3}$, Ta atoms are located in trigonal prisms surrounded by a random mixture of Te and Se atoms. Refining the structure with the ideal 3R atomic coordinates leaves a significant positive residual electron density that is unaccounted for by the model. By investigating the detailed electron density maps, layer stacking faults, which are a common occurrence in crystal structures of this type, were observed through the presence of an "extra" atom site in the tantalum layer, occupied at the 5% level; thus about 5% of the layers in 3R TaSe$_{1.7}$Te$_{0.3}$ crystal studied quantitatively are stacked in a way that violates the ideal A-B-C stacking of the 3R phase (e.g. in an A-B-A-B stacking); the remaining 95% of the structure is unfaulted 3R. The PXRD pattern for 3R-TaSe$_{1.65}$Te$_{0.35}$ is shown in the main panel of Figure 1b. The inset of Figure 1b shows the variation of the room-temperature lattice parameters (*a* and *c*) for 3R-TaSe$_{2-x}$Te$_x$ in its full range of composition stability; the lattice parameters *a* and *c* increase from $a = 3.436(1)$ Å $c = 19.207(9)$ for $x = 0.02$ to $a = 3.465(1)$ Å and $c =$ 19.600(7) Å for x = 0.35. For higher Te contents, a mixture of MX$_2$ structures is encountered until $x$ = 0.8, where TaSe$_{2-x}$Te$_x$ changes into the 1T-polymorph (*P-3m1*, *space group #164*), which exists until $x$ = 1.3. The powder diffraction pattern for one of the 1T compositions is shown in Figure 1c and the inset to Figure 1c shows the variation of the room-temperature lattice parameters (*a* and *c*) for 1T-TaSe$_{2-x}$Te$_x$ over its full range of composition stability. For this phase, the lattice parameters increase linearly from *a* = 3.5468(4) Å *c* = 6.6344(7) Å (x = 0.8) to *a* = 3.6008(2) Å and *c* = 6.5356(10) Å (x = 1.3). A mixture of MX$_2$ phases is encountered again at higher *x* until the distorted 1T structure of TaTe$_2$ is found for $1.8 \leq x \leq 2.0$.

To compare the structural stability regimes of the different forms of TaSe$_{2-x}$Te$_x$, it is most instructive to divide the *c* axis lattice parameter by the number of layers in the stacking repeat, *n*, and then compare the reduced *c/a* ratios (*c/n*)/*a* to define the structural characteristics of single MX$_2$ layers. Figure 1d shows the *x* variation of the reduced *c/a* ratio, (*c/n*)/*a*, for the 2H (*n* = 2, *x* = 0), 3R (*n* = 3, $0.02 \leq x \leq 0.35$) and 1T (*n* = 1, $0.8 \leq x \leq 1.3$) phases. As shown in the plot, the (*c/n*)/*a* ratio increases with increasing *x* in the 3R form, with a at the 2H to 3R transition, and is always less than the ideal ratio, which for the close-packed trigonal-prismatic arrangement is considered to be 2.0 (26). The (*c/n*)/*a* ratio collapses for the 1T polymorph, and changes relatively little with increasing *x*. In this case, the (*c/n*)/*a* ratios are slightly larger than the ideal value of 1.633 (26). The (*c/n*)/*a* ratio where the 3R polytype becomes unstable and the 1T polymorph becomes stable is consistent with expectations for MX$_2$ phases, as has been described by others (26).

To determine whether CDWs are present in the 3R and 1T forms of TaSe$_{2-x}$Te$_x$, the materials were studied at low temperatures by electron diffraction and scanning tunneling microscopy (STM). The temperature dependent electron diffraction patterns obtained from single-crystal domains (Figures 2a and 2b) reveal critical information about the CDWs in both materials. In the 3R form, a strong CDW appears on cooling TaSe$_{2-x}$Te$_x$ below ambient temperature. The CDW gives rise to extra peaks in the electron diffraction patterns, which are already weakly visible at 330 K. For 3R TaSe$_{1.9}$Te$_{0.1}$, the superlattice peaks are first in incommensurate positions and weak but then sharpen and intensify significantly on reducing temperature until at 10 K they are sharp and strong, indicating that the CDW is well defined and ordered over a long range at low temperatures. The incommensurate locations of the spots in reciprocal space at higher temperatures shows that first there is an ICDW phase with the *q* vector larger than 0.33. The ICDW diffraction spots then change position on cooling until they lock into nearly commensurate positions, $q \sim 0.32$, at around 100 K. The illuminated areas from which the electron diffraction patterns were obtained is relatively large, with beam diameters greater than 300 nm. The small value of incommensuration observed in the low temperature locked-in CDW phase, which is less than 0.01 from the commensurate value of 0.33, may come from defects and domain walls in the low temperature CDW phase; as the STM topographic images show (see below) the CDW is locked into a commensurate relationship with the underlying atomic lattice over the vast majority of the material, we thus consider it to be a commensurate CDW with a wave vector of q = 0.33. The 10 K electron diffraction pattern shown in Figure 2a is an indication of the quality of this low temperature CCDW phase. At the lock-in transition, the intensity of the diffracted spots from the CDW increases dramatically, an indication of its strengthening in the CCDW state. The temperature dependent characteristics of the CDW in 3R TaSe$_{1.9}$Te$_{0.1}$ obtained from the electron diffraction study are summarized in Figure 2c. Thus, in analogy to the 2H form of TaSe$_2$, 3R-TaSe$_{1.9}$Te$_{0.1}$ first has an ICDW and then locks in to a CCDW phase on further cooling. The $q$ vectors of the CCDW phase, $q_1 = q_2 = 0.33$, indicate a tripling of the in-plane unit cell along both $a_1$ and $a_2$ by the CDW.

STM measurements on 3R-TaSe$_{1.9}$Te$_{0.1}$ and on 3R-TaSe$_{1.7}$Te$_{0.3}$ provide additional characterization of the CCDW phase. Topographic images on the atomic scale (Fig. 3) display the in-plane unit cell tripling in real space on the surface of both samples below the CDW transition temperature. While the CDW superlattice in TaSe$_{1.9}$Te$_{0.1}$ is clearly visible (Fig. 3a-b), that in TaSe$_{1.7}$Te$_{0.3}$ is moderately masked by the disorder (Fig. 3d-e). However, the Fourier transform of the topographic images for both samples (Fig. 3c and 3f) unambiguously reveals the primary peaks of the CCDW at $q_1 = q_2 = 0.33$. Hence, the CDW is present for both high and low Te contents in the 3R phase. Remarkably, even in the presence of the strong disorder we observe that the phase of the CDW is unperturbed and only a single domain CDW appears in the field of view (~40 nm x 40 nm). Furthermore, the STM data indicates that the apparent tripling of the cell by the CDW in both in-plane directions deduced from the electron diffraction patterns is not due to the overlap of single-$q$ domains in different orientations; i.e. it is a 2D CDW. Further, the diffracted spots from the CDW phase are visible in single crystal diffraction experiments at 100 K; the data shows that the $q$ vector is in-plane only; there is no $c$ axis component. The data therefore show that for the CCDW in the 3R polymorph, $q_1$ = 0.33, $q_2$ = 0.33, and $q_3$ = 0). The CCDW state the 2H-TaSe$_2$ polytype is also reported have the wave-vectors $q_1$ = $q_2$ = 0.33 and $q_3$ = 0 (27,28). In other words, in both 2H and 3R polytypes of TaSe$_2$, the electronic instabilities that lead to the CCDWs are two-dimensional in character.

The situation is somewhat different for the 1T polymorph (Figures 2b and 2d). In this case, weak, diffuse superlattice spots whose intensities and positions are relatively independent of temperature between 10 and 330 K are observed in the electron diffraction experiments. Here the in-plane $q$ vector is further from the commensurate value, near q = 0.30, but the weakness and diffuseness of the spots makes them invisible in a single crystal diffraction experiment and so we have no information about their *c* axis component. These spots likely represent an ICDW phase that is ordered over short spatial distances and stable over the whole temperature range studied. Alternatively they may have a chemical origin, such as might occur due to short range Se/Te ordering. Further work will be required to determine which is the case.

We next consider the electronic properties of the phases. The main panel of Figure 4a shows the temperature dependence of the normalized electrical resistivity, ($\rho/\rho_{300K}$), for the polycrystalline 3R-TaSe$_{2-x}$Te$_x$ (0.02 ≤ x ≤ 0.35) samples. All the 3R samples have resistivities below 10 mohm cm at 300 K with a metallic temperature dependence, and all show the signature of the lock-in to the CCDW phase at around 100 K through a change in slope of $\rho$(T). A similar change in slope of $\rho$(T) is observed in many TMDC systems at the onset of a CDW that localizes some but not all of the electrons at the Fermi surface (29). A look at the derivative of the $\rho$(T) curves (inset of Figure. 4d) indicates that the impact of the CDW lock-in transition, which appears to have a temperature that is independent of Te content, weakens with increasing Te content in the 3R phase. Correspondingly, a superconducting transition is found (inset of Figure 4a). With higher Te doping in the 3R phase, the superconducting transition temperature (T$_c$) increases. The maximum superconducting transition temperature is 2.4 K for 3R-TaSe$_{1.65}$Te$_{0.35}$. T$_c$ is a factor of approximately 20 higher than is observed in the 2H form. The superconducting transition is clearly observed through the presence of a full shielding signal in the temperature dependent magnetization measurements (Figure 4d).

The main panel of Figure 4b shows the temperature dependence of the normalized electrical resistivity ($\rho/\rho_{300K}$) for the polycrystalline samples of 1T-TaSe$_{2-x}$Te$_x$ ($0.8 \leq x \leq 1.3$). In this case, the residual-resistivity-ratio is very small, RRR = $\rho_{300K}/\rho_n < 1.3$, which we take as a reflection of the substantial Se-Te disorder present. No signature of a CDW lock-in transition is seen in $\rho(T)$, consistent with the electron diffraction data. At low temperatures, a clear, sharp ($\Delta T_c < 0.1$ K) drop of $\rho(T)$ is observed, signifying the onset of superconductivity (inset of Figure 4b). The sample with x = 1 shows the highest $T_c$, 0.73 K. Finally, Figure 3c presents the temperature dependence of the normalized resistivity for polycrystalline samples of the monoclinic polymorph of TaSe$_{2-x}$Te$_x$ ($1.8 \leq x \leq 2$). This variant shows metallic behavior, similar to what has been previously reported (30), with no superconducting transition down to 0.4 K.

Further information on the electronic properties and superconductivity in the 3R and 1T variants of TaSe$_{2-x}$Te$_x$ was obtained from specific heat measurements. The main panels of Figure 4e and f show the temperature dependence of the zero-field specific heat, $C_p/T$ versus $T^2$, for 3R-TaSe$_{1.65}$Te$_{0.35}$ and 1T-TaSe$_{1.2}$Te$_{0.8}$. The normal state specific heat at low temperatures (but above Tc) obeys the relation of $C_p = \gamma T + \beta T^3$, where $\gamma$ and $\beta$ describe the electronic and phonon contributions to the heat capacity, respectively, the latter of which is a measure of the Debye Temperature ($\theta_D$). By fitting the data in the temperature range of 2 - 10 K, we obtain the electronic specific heat coefficients $\gamma = 7.25$ mJ·mol$^{-1}$·K$^{-2}$ for 3R TaSe$_{1.65}$Te$_{0.35}$ and $\gamma = 2.91$ mJ·mol$^{-1}$·K$^{-2}$ for 1T-TaSe$_{1.2}$Te$_{0.8}$, and the phonon specific heat coefficients $\beta = 0.93$ mJ·mol$^{-1}$·K$^{-4}$ for 3R-TaSe$_{1.65}$Te$_{0.35}$ and $\beta = 1.82$ mJ·mol$^{-1}$·K$^{-4}$ for 1T-TaSe$_{1.2}$Te$_{0.8}$. Using these values of $\beta$, we estimate the Debye temperatures by the relation $\theta_D = (12\pi^4 nR/5\beta)^{1/3}$, where n is the number of atoms per formula unit (n = 3), and R is the gas constant; the $\theta_D$ values are found to be 184 K for 3R-TaSe$_{1.65}$Te$_{0.35}$ and 147 K for 1T-TaSe$_{1.2}$Te$_{0.8}$. As shown in the insets for Figures 3e and f, both materials display a large specific heat jump at $T_c$. The superconducting transition temperatures are in excellent agreement with the $T_c$s determined in the $\rho(T)$ measurements. We estimate $\Delta C/T_c = 8.7$ mJ mol$^{-1}$ K$^{-2}$ for 3R-TaSe$_{1.65}$Te$_{0.35}$ and $\Delta C/T_c = 3.93$ mJ mol$^{-1}$ K$^{-2}$ for 1T-TaSe$_{1.2}$Te$_{0.8}$. The normalized specific heat jump values $\Delta C/\gamma T_c$ are found to be 1.20 for 3R-TaSe$_{1.65}$Te$_{0.35}$ and 1.35 1T-TaSe$_{1.2}$Te$_{0.8}$, which are near that of the Bardeen-Cooper-Schrieffer (BCS) weak-coupling limit value (1.43), confirming bulk superconductivity. Using the Debye temperature ($\theta_D$), the critical temperature $T_c$, and assuming that the electron-phonon coupling constant ($\lambda_{ep}$) can be calculated from the inverted McMillan formula (31):

$$\lambda_{ep} = \frac{1.04 + \mu^* \ln\left(\frac{\theta_D}{1.45 T_C}\right)}{(1 - 0.62\mu^*)\ln\left(\frac{\theta_D}{1.45 T_C}\right) - 1.04}.$$

the values of $\lambda_{ep}$ obtained are 0.64 for 3R-TaSe$_{1.65}$Te$_{0.35}$ and 0.51 for 1T-TaSe$_{1.2}$Te$_{0.8}$ and suggest weak coupling superconductivity. With the Sommerfeld parameter ($\gamma$) and the electron-phonon coupling ($\lambda_{ep}$), the density of states at the Fermi level can be calculated from $N(E_F) = \frac{3}{\pi^2 k_B^2 (1 + \lambda_{ep})}\gamma$. This yields $N(E_F)$ = 1.88 states/eV f.u. for optimal 3R-TaSe$_{1.65}$Te$_{0.35}$ and $N(E_F)$ = 0.82 states/eV f.u. for 1T-TaSe$_{1.2}$Te$_{0.8}$. These compare to $\lambda_{ep}$ = 0.4 and $N(E_F)$ = 1.55 for the 2H-TaSe$_2$ (32). The somewhat larger $N(E_F)$ and $\lambda_{ep}$ values for the 3R polytype may be why it has dramatically higher Tc than the 2H polytype, but why these parameters are different in the 2H and 3R polytypes is not currently known.

Finally, the overall behavior of the TaTe$_x$Se$_{2-x}$ system is summarized in the structural and electronic phase diagram shown in Figure 5. On Te substitution for Se in 2H-TaSe$_2$, the 3R polytype is immediately stabilized and in TaSe$_{2-x}$Te$_x$ exists in the range of 0.02 ≤ x ≤ 0.35. 3R TaSe$_{2-x}$Te$_x$ shows the coexistence of a CDW and superconductivity in this composition range. The superconducting transition temperature is found to maximize at the limit of the structural stability of the 3R phase, $x$ = 0.35. The maximum $T_c$, 2.4 K, is 17 times higher than that for 2H-TaSe$_2$. We conclude that 3R-TaSe$_{2-x}$Te$_x$ can be considered as a good candidate for characterizing the balance between CDW formation and superconductivity in the 3R polytype of the layered TMDCs. With further Te doping, the 3R-polytype becomes unstable as its ($c/n$)/$a$ ratio approaches the structural stability limit expected for TMSCs, and the 1T polymorph, with octahedral rather than trigonal prismatic coordination for the Ta, emerges at $x$ = 0.8. The 1T polytype structure exists from $x$ = 0.8 to 1.3. 1T-TaSe$_{2-x}$Te$_{x-x}$ displays superconducting transitions below 1 K and the $T_c$ does not change significanly with $x$. 1T-TaSe$_{2-x}$Te$_x$ appears to display a weak, short range ordered incommensurate CDW at temperatures as high as 330 K, but further work will be necessary to support that conclusion. The monoclinic polymorph exists over a limited composition range in TaSe$_{2-x}$Te$_x$, from $x = 1.8$ to 2.0; it is metallic but not superconducting above 0.4 K.

In conclusion we have shown that the TaSe$_{2-x}$Te$_x$ system, based on the isoelectronic substitution of Te for Se in TaSe$_2$, is an excellent venue for investigating the influence of polytypism and polymorphism on superconductivity in the layered transition metal dichalcogenides. It may be that the major impact of the change in polytype from 2H to 3R in this system, which increases Tc by a factor of 17, is in the the final analysis actually still a 2D effect. The 3R polytype is stable for larger (c/n)/a ratios for a single layer than it is possible to obtain in the 2H polytype: (c/n)/a for the 2H polytype is 1.84 whereas for the maximum Tc of the 3R polytype it is 1.89. If this is the primary reason for the difference in Tc, then it indicates a remarkable sensitivity of Tc to the aspect ratio of the TaX$_6$ triangular prisms that make up the single layers. Alternatively, the superconducting transition temperature may depend on differences in the electronic structure that arise as a result of the differences between a two layer stacking repeat and a three layer stacking repeat, in other words how the nominally 2D Fermi surface is modulated along *c*, the perpendicular direction, by the stacking. Further investigation will be required to resolve which of these is the case, or whether a different strong influence on Tc is present. The Tc of the 1T polymorph is intermediate between that of the 2H and the 3R, with a Tc a factor of 5 larger than that in the 2H variant at approximately the same (c/n)/a, but as the TaX$_6$ coordination scheme is octahedral rather than trigonal prismatic in this polymorph, different aspects of its electronic structure may determine its superconducting transition temperature. Further study of this system may prove to be of significant interest.

**Methods**

Polycrystalline samples of TaSe$_{2-x}$Te$_x$ ($0 \leq x \leq 2$) were synthesized in two steps by solid state reaction. First, mixtures of high-purity fine powders of Ta (99.8%), Te (99.999%) and Se (99.999%) in the appropriate stoichiometric ratios were thoroughly ground, pelletized and heated in sealed evacuated silica tubes at a rate of 1 $^o$C/min to 700 $^o$C and held there for 48 h. Subsequently, the as-prepared powders were reground, re-pelletized and sintered again, heated at a rate of 3 $^o$C/min to 1000 $^o$C and held there for 48 h. Single crystals were grown by chemical vapor transport (CVT) method with iodine as a transport agent. 1 gram as-prepared powders TaSe$_{2-x}$Te$_x$ mixed with 50 mg iodine were sealed in sealed evacuated silica tubes and heated for one weeks in a two zone furnace, where the temperature of spurce and growth zones were fixed at 1050 °C and 1000 °C, respectively. The identity and phase purity of the samples was determined by powder X-ray diffraction (PXRD, Rigaku, Cu Kα radiation, graphite diffracted beam monochromator). Unit cell parameters were refined from the powder diffraction data through use of the FULLPROF diffraction suite (33). Measurements of the temperature dependence of the electrical resistivity were heat capacity performed in a Quantum Design Quantum Design Physical Property Measurement (PPMS). For the superconducting samples, $T_c$ was taken as the intersection of the extrapolation of the steepest slope of the resistivity $\rho(T)$ in the superconducting transition region and the extrapolation of the normal state resistivity ($\rho_n$) (34). Magnetization measurements as a function of temperature and applied field were carried out in a Quantum Design Magnetic Property Measurement (MPMS). Selected resistivities for $TaSe_{2-x}Te_x$ ($0.1 \leq x \leq 0.25$, $0.7 \leq x \leq 1.3$ and 1.8, 1.9, 2), and heat capacities for $TaSe_{1.65}Te_{0.35}$ and $TaSe_{1.2}Te_{0.8}$ were measured in the PPMS equipped with a $^3$He cryostat.

Single crystals from the samples were mounted on the tips of glass fibers. Room temperature intensity data were collected on a Bruker Apex II diffractometer with Mo radiation $Ka_1$ ($\lambda=0.71073$ Å). Data were collected over a full sphere of reciprocal space with 0.5° scans in ω with an exposure time of 10s per frame. The 2θ range extended from 4° to 60°. The SMART software was used for data acquisition. Intensities were extracted and corrected for Lorentz and polarization effects with the SAINT program. Empirical absorption corrections were accomplished with SADABS which is based on modeling a transmission surface by spherical harmonics employing equivalent reflections with $I > 2\sigma(I)$ (35-38). With the SHELXTL package, the crystal structures were solved using direct methods and refined by full-matrix least-squares on $F^2$ (37). All crystal structure drawings were produced using the program *VESTA* (38).

Prior to the STM measurements, the samples were cleaved and transported to the microscope in ultra-high vacuum. The experiments were performed on $TaSe_{1.9}Te_{0.1}$ at 48 K and on $TaSe_{1.7}Te_{0.3}$ at 27 K with a home-built variable temperature STM. Temperature-dependent electron diffraction measurements were performed on a JEOL 2100F microscope at Brookhaven National Laboratory equipped with a liquid-helium cooling sample holder.

**Acknowledgements**

This research was primarily supported by the ARO MURI on superconductivity, grant FA-9550-09-1-0953. ARO MURI grant FA-9550-10-1-0553 supported the single crystal diffraction work, and the DOE supported the powder diffraction work of J. K. through grant DE FG02-08ER46544. The electron diffraction study at Brookhaven National Laboratory was supported by the DOE BES, by the Materials Sciences and Engineering Division under contract DE-AC02-98CH10886, and through the use of the Center for Functional Nanomaterials. STM work at Princeton was supported under DOE-BES, ARO-MURI program W911NF-12-1-0461, and NSF-DMR-1104612.

**Author Contributions**

R.J.C. and H.X.L conceived and designed the experiments and H.X.L performed the synthetic experiments. H.X.L. and R.J.C. analyzed and interpreted the data. J.T and Y.M.Z performed electron diffraction experiments and analysis. J.K. and H.X.L. performed and analyzed the XRD refinement data. W.W.X performed the single diffraction and analyzed the structural data. H.I, A.G. and A.Y. performed the STM experiments and analysis. H.X.L and R.J.C were the primary authors of the paper, with contributions to the text from all authors. All authors approved the content of the manuscript.

**Competing financial interests:** The authors declare no competing financial interest.

**References:**

1. Withers R. L, Bursil L. A (1981) The structure of the incommensurate superlattices of 2H-TaSe$_2$, *Philosophical Magnzine B*43 (4):635-672.
2. Arguello C. J., Chockalingam S. P., Rosenthal E. P., Zhao L., Gutiérrez C., Kang J. H., Chung W. C., Fernandes R. M., Jia S., Millis A. J., Cava R. J., Pasupathy A. N (2014), Visualizing the charge density wave transition in 2H-NbSe$_2$ in real space, *Phys. Rev. B* 89(23):235115-235123.
3. Soumyanarayanana A., Yee Michael M., He Y., van Wezel J., Rahne Dirk J., Rossnagel K., Hudson E. W., Norman Michael R., and Hoffman J. E (2013) Quantum phase transition from triangular to stripe charge order in NbSe$_2$, *Proc Natl Acad Sci USA* 110(10):1623-1627.
4. Malliakas C. D, Kanatzidis M. G (2013) Nb-Nb Interactions define the charge density wave structure of 2H-NbSe$_2$. *J Am Chem Soc* 135(5):1719-1722.
5. Grüner G (1988) The dynamics of charge-density waves, *Rev Mod Phys* 60(4):1129-1181.
6. Sipos B, Kusmartseva A. F, Akrap A, Berger H, Forro L, Tutis E (2008) From Mott state to superconductivity in 1T-TaS$_2$. *Nat Mater* 7:960-965.
7. Wilson J. A, Yoffe A D (1969) The transition metal dichalcogenides discussion and interpretation of the observed optical, electrical and structural properties. *Adv Phys* 18(73):193-335.
8. Yokota K, Kurata G, Matsui T, Fukuyama H (2000) Superconductivity in the quasi-two-dimensional conductor 2H-TaSe$_2$. *Phys B* 284-288:551-552.
9. Rossnagel K, Smith N.V (2007) Spin-orbit splitting, Fermi surface topology, and charge-density-wave gapping in 2H-TaSe$_2$. *Phys. Rev. B* 76:073102.
10. Wilson J. A (1978) Questions concerning the form taken by the charge-density wave and the accompanying periodic-structural distortions in 2H-TaSe$_2$, and closely related materials. *Phys. Rev. B* 17:3880-3896.
11. Ruzicka B, Degiorgi L (2001) Charge Dynamics of 2H-TaSe$_2$ along the Less-Conducting c-Axis. *Phys. Rev. Lett* 86:4136.
12. Sugai S, Murase K (1982). Generelized electronic susceptibility and charge density waves in 2H-TaSe$_2$ by Raman scattering. *Phys. Rev. B* 25:2418.

13. Rossangel K (2011) On the top origin of charge density waves in select layerd transition metal dichlcogenides. *J Phys. Condens mater* 23:213001-213024.

14. Nakashizu T, Sekine T, Uchinokura K, Matsuura E (1984) Raman study of charge-density-wave excitations in 4Hb-TaS$_2$. *Phys. Rev. B* 29:3090.

15. Hajiyev P, Cong C X, Qiu C Y, Yu T (2013) Contrast and Raman spectroscopy study of single- and few-layered charge density wave material: 2H-TaSe$_2$. *Sci Rep* 3:2593.

16. Tsang J.; Smith J.; Shafer M.; Meyer S (1977). Raman spectroscopy of the charge-density-wave state in 1T- and 2H-TaSe$_2$. *Phys. Rev. B* 16:4239.

17. Inosov D, Evtushinsky D, Zabolotnyy V, Kordyuk A, Büchner B, Follath R, Berger H, Borisenko S (2009) Temperature-dependent Fermi surface of 2H-TaSe$_2$ driven by competing density wave order fluctuations. *Phys. Rev. B* 79:125112.

18. Moncton D, Axe J, DiSalvo F (1975) Study of superlattice formation in 2H-NbSe$_2$ and 2H-TaSe$_2$ by neutron scattering. *Phys Rev Lett* 34:734.

19. Huisman R, Jellinek F (1969) On the polymorphism of tantalum diselenide, *J Less Common Met* 17:111.

20. Brown B. E, Beerntsen D. J (1965) Preparation and Crystal Growth of Materials with Layered Structures. *Acta Crystallogr.* 18:31-36.

21. Lieth R. M. A. (Ed.) (1977) Physics and Chemistry of Materials with Layered Structures, Vol. 1, D. Reidel Publishing Company: Dordrecht, Holland,

22. Bjerkelund E, Kjkshus A (1967) On the structural properties of the Ta$_{1+x}$Se$_2$ phase. *Acta Chem Scand*, 21:513-526.

23. Liu Y, Ang R, Lu W J, Song W H., Li L J., Sun Y P (2013) Superconductivity induced by Se-doping in layered charge-density-wave system 1T-TaS$_{2-x}$Se$_x$. *Appl. Phys. Lett* 102:192602.

24. Vernes A, Ebert H, Bensch W, Heid W, Näher C (1998) Crystal structure, electrical properties and electronic band structure of tantalum ditelluride. *J. Phys.:Condens. Matter* 10:761

25. Huisman R, Kadijk F., Jellinek F (1970) The non-stoichiometric phases Nb$_{1+x}$Se$_2$ and Ta$_{1+x}$Se$_2$. *J Less Common Met*, 21:187-193.

26. Kertesz M, Hoffmann R (1984) Octahedral vs. Trigonal-Prismatic Coordination and Clustering in Transition-Metal Dichalcogenides, *J Am Chem Soc* 106, 3453-3460.

27. Moncton, D. E.; Axe, J. D.; DiSalvo, F. J. (1977) Neutron scattering study of the charge-density wave transitions in 2H-TaSe$_2$ and 2H-NbSe$_2$. *Phys Rev B*, 16:801.

28. Jericho M H, Ott H R, Rice T M (1986) Effect of discommensurations on the electrical resistivity of 2H-TaSe$_2$. *J. Phys. C: Solid State Phys* 9:1377-1387.

29. Wilson J. A, Di Salvo F. J, Mahajan S (1974). Charge-density waves in metallic, layered, transition-metal dichalcogenides. *Phys Rev Lett* 32:882.

30. Vernesyx A, Eberty H, Benschz W, Heidz W, Näther C (1998) Crystal structure, electrical properties and electronic band structure of tantalum ditelluride. *J. Phys: Condens Matter* 10:761-774.

31. McMillan W L (1968) Transition temperature of strong-coupled superconductors. *Phys. Rev*. 167:331.

32. Subba Rao G. V, Shafer M. W. Intercalation in layered transition metal dichalcogenides, Book, P116.

33. Juan. Rodŕguez-Carvajal (2001) Recent developments of the program FULLPROF. *Comm. Powder Diffr*. 26:12-19.

34. Klimczuk T, Cava R. J (2004) Carbon isotope effect in superconducting MgCNi$_3$. *Phys. Rev. B* 70:212514.

35. Sheldrick, G. M. (2001) .*SADABS*, University of Gottingen, Gottingen, Germany.

36. Sheldrick, G. M. (2008) A short history of SHELX. *Acta Crystallogr. A* 64: 112.

37. *SHELXTL*, version 6.10, Bruker AXS Inc.: Madison, WI, 2000.

38. Momma K, Izumi F (2011) VESTA 3 for three-dimensional visualization of crystal, volumetric and morphology data. *J. Appl. Crystallogr* 44:1272.

39. LCvy F, Froidevaux Y (1979) Structural and electrical properties of layered transition metal selenides V$_x$Ti$_{1-x}$Se$_2$ and Ta$_x$Ti$_{1-x}$Se$_2$. *J. Phys. C: Solid State Phys*. 12:473-487.

**Table 1. Single crystal crystallographic data for 3R TaSe$_{1.70}$Te$_{0.30}$ at 296(2) K.**

| | |
|---|---|
| Formula | TaSe$_{1.70}$Te$_{0.30}$ |
| F.W. (g/mol); | 353.46 |
| Space group; $Z$ | $R3m$(No.160); 3 |
| $a$ (Å) | 3.4603(7) |
| $c$ (Å) | 19.523(4) |
| $V$ (Å$^3$) | 202.44(9) |
| Absorption Correction | Multi-Scan |
| Extinction Coefficient | 0.006(3) |
| $\mu$(mm$^{-1}$) | 66.441 |
| $\theta$ range (deg) | 3.130-29.555 |
| $hkl$ ranges | $-4 \leq h,k \leq 4$ <br> $-27 \leq l \leq 27$ |
| No. reflections; $R_{int}$ | 714; 0.0451 |
| No. independent reflections | 189 |
| No. parameters | 14 |
| $R_1$; $wR_2$ (all $I$) | 0.0530; 0.1347 |
| Goodness of fit | 1.284 |
| Diffraction peak and hole (e$^-$/Å$^3$) | 7.885; −3.371 |

**Table2** Atomic coordinates and equivalent isotropic displacement parameters of TaSe$_{1.70}$Te$_{0.30}$. U$_{eq}$ is defined as one-third of the trace of the orthogonalized U$_{ij}$ tensor (Å$^2$).

| Atom | Wyckoff. | Occupancy. | $x$ | $y$ | $z$ | $U_{eq}$ |
|---|---|---|---|---|---|---|
| Ta1 | 3$a$ | 0.95(1) | 0 | 0 | 0.2931(1) | 0.009(1) |
| Ta2 | 3$a$ | 0.05(1) | 2/3 | 1/3 | 0.293(3) | 0.009(1) |
| Se/Te1 | 3$a$ | 0.85/0.15 | 1/3 | 2/3 | 0.2059(3) | 0.012(1) |
| Se/Te2 | 3$a$ | 0.85/0.15 | 1/3 | 2/3 | 0.3804(3) | 0.011(1) |

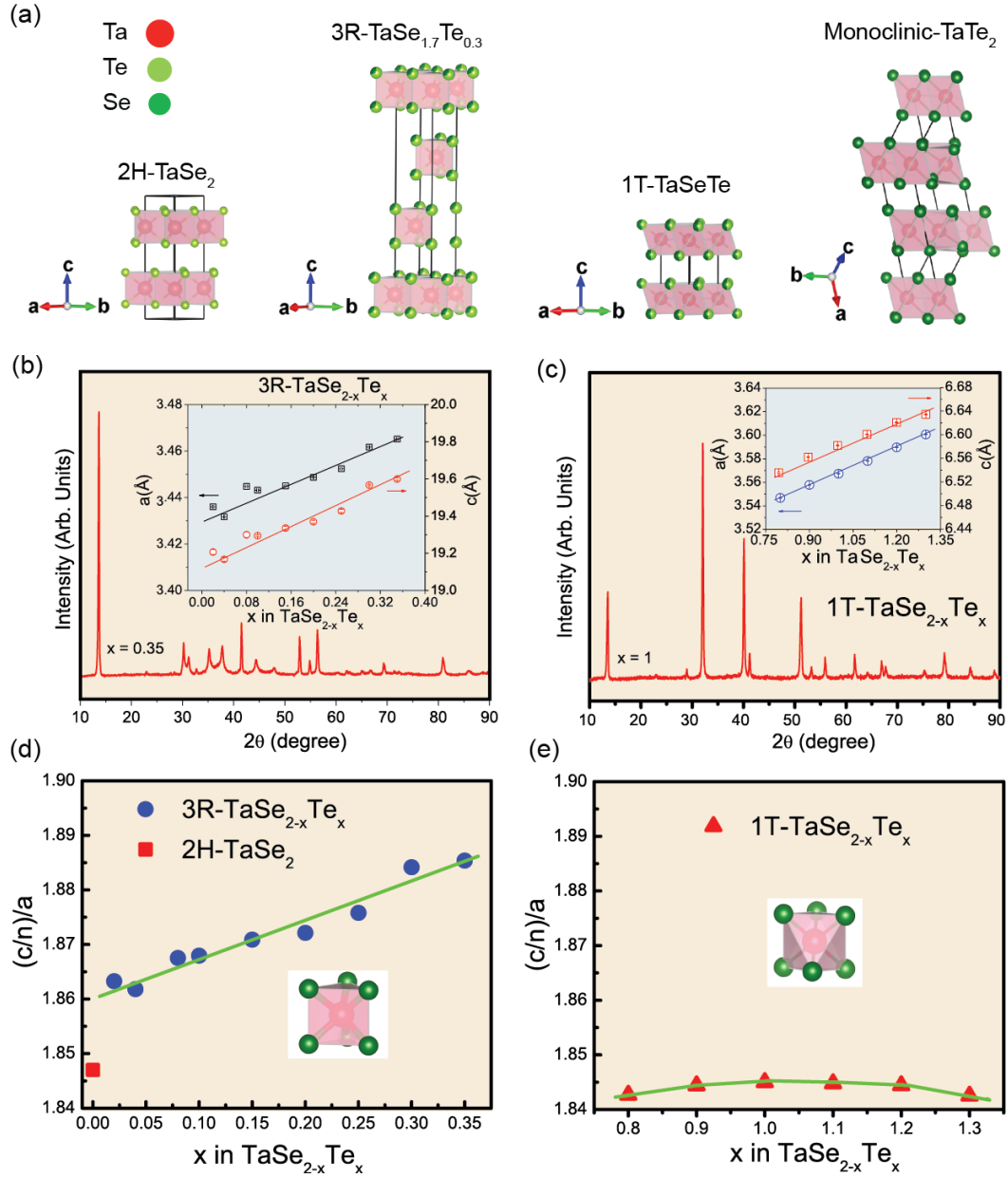

**Figure 1. Structural characterization and analysis of the polytypes and polymorphs of TaSe$_{2-x}$Te$_x$.** (a) Crystal structures of 2H-TaSe$_2$, 3R-TaSe$_{1.65}$Te$_{0.35}$, 1T-TaSeTe, and monoclinic TaTe$_2$. (b) Powder X-ray diffraction pattern for 3R-TaSe$_{1.65}$Te$_{0.35}$. Inset: lattice parameters for 3R-TaSe$_{2-x}$Te$_x$ (0.02 ≤ x ≤ 0.35), (c) powder X-ray diffraction pattern for 1T-TaTeSe, inset: lattice parameters for 1T-TaSe$_{2-x}$Te$_x$ (0.8 ≤ x ≤ 1.3), (d) and (e) The reduced lattice parameter ratio, (c/n)/a, for 2H, 3R, and 1T TaSe$_{2-x}$Te$_x$. (c/n)/a data for 2H-TaSe$_2$ is from Ref.[39]

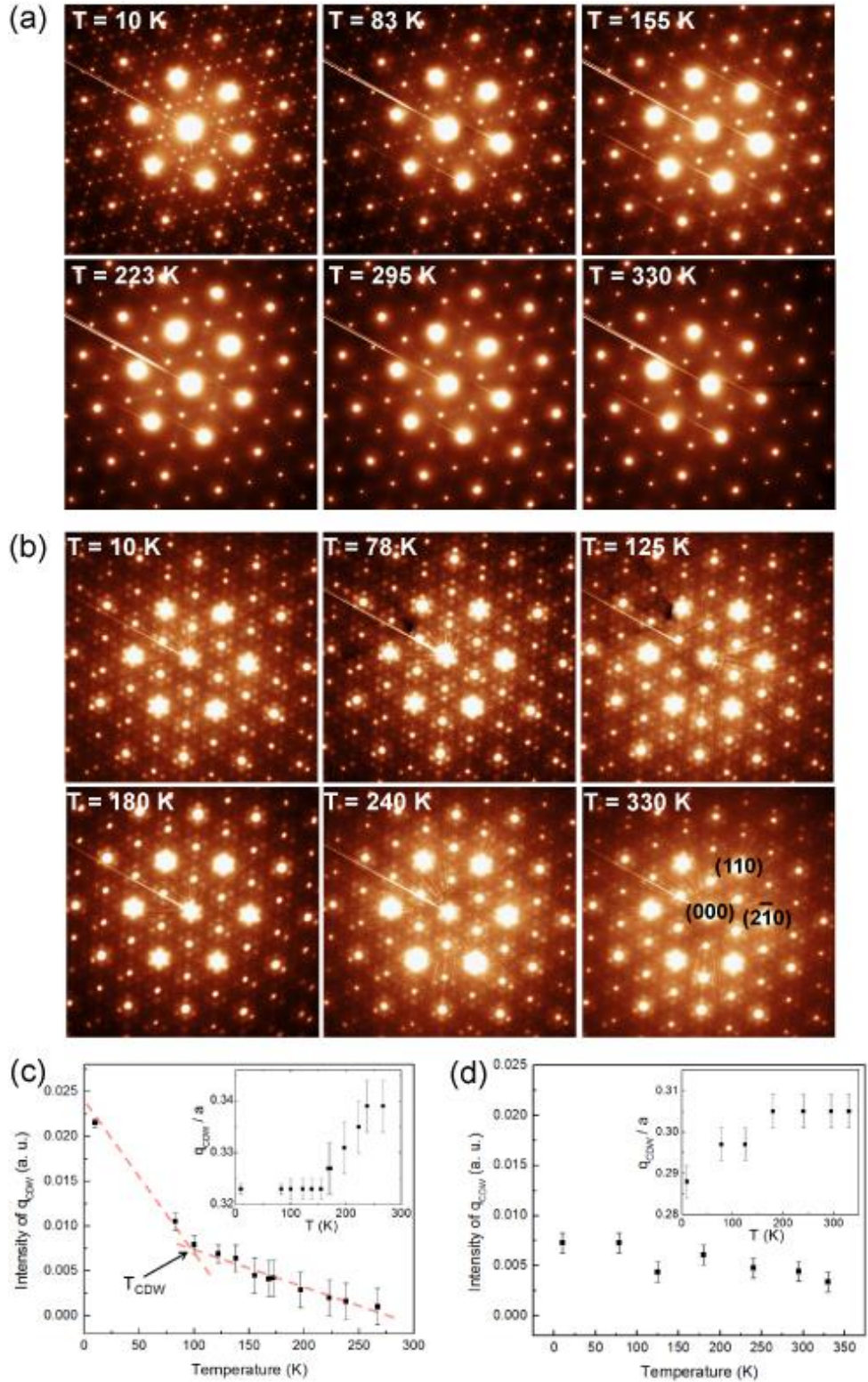

**Figure 2. Characterization of the Charge Density waves in the 3R and 1T polymorphs.** Temperature dependence of the incommensurate CDW state in a-b plane (a) temperature dependence of electron diffraction patterns of polycrystalline 3R-TaSe$_{1.7}$Te$_{0.3}$, (b) temperature dependence of electron diffraction patterns of polycrystalline 1T-TaTeSe; CDW wave vector q$_{CDW}$ as a function of temperature for (c) 3R-TaSe$_{1.6}$Te$_{0.3}$ and (d) 1T-TaSeTe.

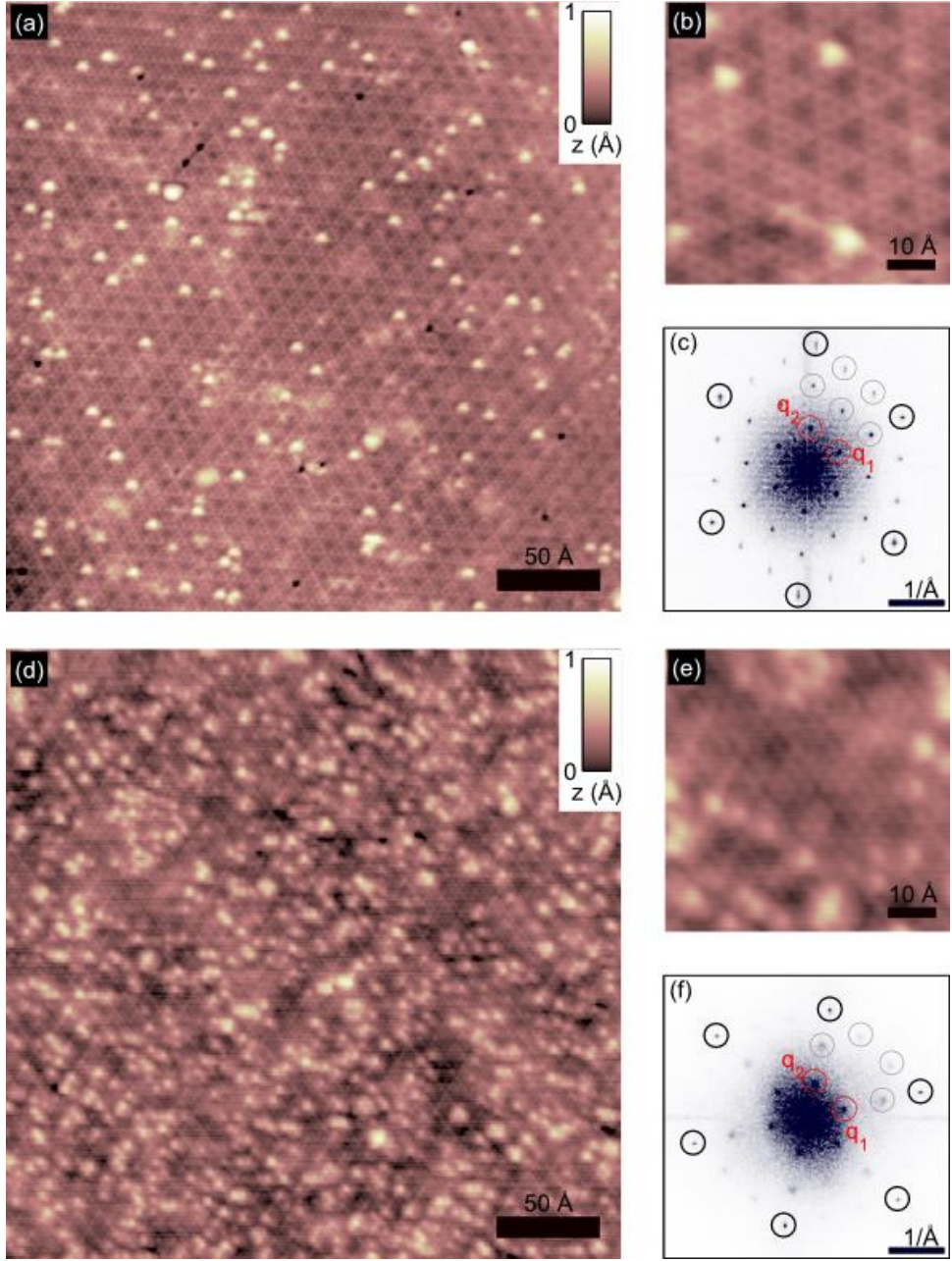

**Figure 3. Visualization of the charge density wave on the surface of 3R-TaSe$_{1.9}$Te$_{0.1}$ (a-c) and 3R-TaSe$_{1.7}$Te$_{0.3}$ (d-f) by STM.** Real space topographic images of (a) 300 Å x 300 Å and (b) 60 Å x 60 Å areas on the cleaved *a-b* surface of TaSe$_{1.9}$Te$_{0.1}$ ($V_{Bias}$ = - 800 mV, (a) $I$ = 100 pA and (b) $I$ = 60 pA) at $T$ = 48 K, which show the tripling of the in-plane unit cell. The CDW remains unchanged around the bright spots on the surface, which are associated with the substituted Te atoms. (c) The Fourier transform of 440 Å x 440 Å large topographic image reveals wave vectors corresponding to the atomic modulation (black circles) and *q* vectors of the commensurate charge density wave phase ($q_1 = q_2 = 0.33$ - red circles). Higher harmonics of $q_1$ and $q_2$ are marked by gray circles. Similar topographic images (d-e) of the surface of TaSe$_{1.7}$Te$_{0.3}$ at $V_{Bias}$ = 300 mV, $I$ =200 pA and $T$ = 27K, and (f) Fourier transform of a 490 Å x 490 Å large area. The CDW is clearly observed in spite of the disorder induced by the higher Te substitution.

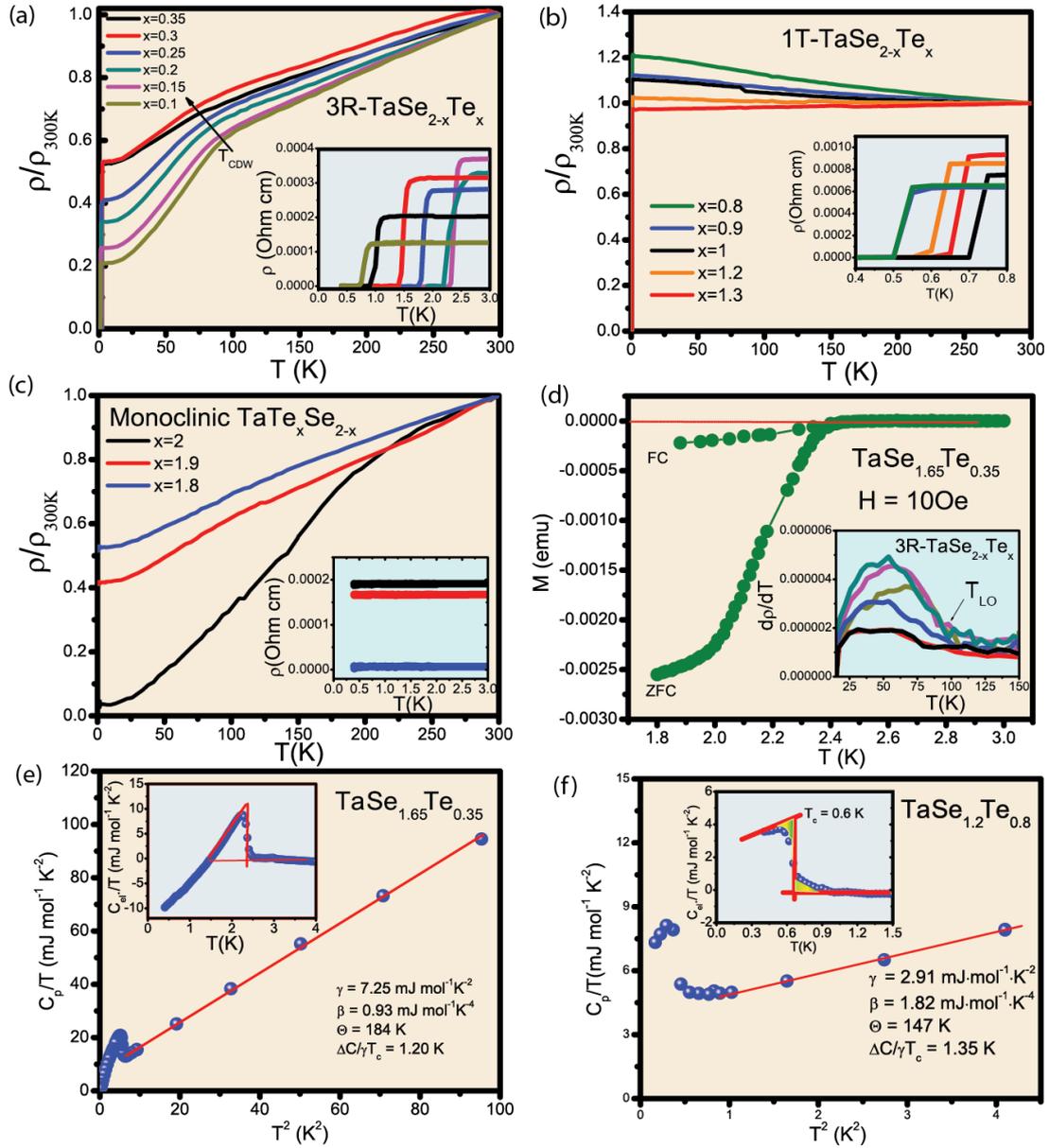

**Figure 4. Characterization of the electronic properties of TaSe$_{2-x}$Te$_x$.** The temperature dependence of the ratio ($\rho/\rho_{300K}$) for (a) 3R-TaSe$_{2-x}$Te$_x$, Inset: enlarged view of low temperature region (0.4 - 3 K), showing the superconducting transition. (b) 1T-TaSe$_{2-x}$Te$_x$, Inset: enlarged view of low temperature region (0.4 – 0.8 K) showing the superconducting transition; (c) Monoclinic-TaSe$_{2-x}$Te$_x$ (1.8 ≤ x ≤ 2), Inset, enlarged view of the low temperature region showing the absence of superconductivity.(d) The temperature dependence of dc magnetic susceptibility for 3R-TaSe$_{1.65}$Te$_{0.35}$ Inset: enlarged view of d$\rho$/dT of 3R-TaSe$_{2-x}$Te$_x$ showing CDW lock in temperature (T$_{LO}$) (e) The temperature dependence of specific heat C$_p$ of 3R-TaSe$_{1.65}$Te$_{0.35}$, presented in the form of C$_p$/T vs T$^2$ (main panel) and C$_{el}$/T vs T (inset). (f) The temperature dependence of specific heat C$_p$ of 1T-TaSe$_{1.2}$Te$_{0.8}$, presented in the form of C$_p$/T vs T$^2$ (main panel) and C$_{el}$/T vs T (inset).

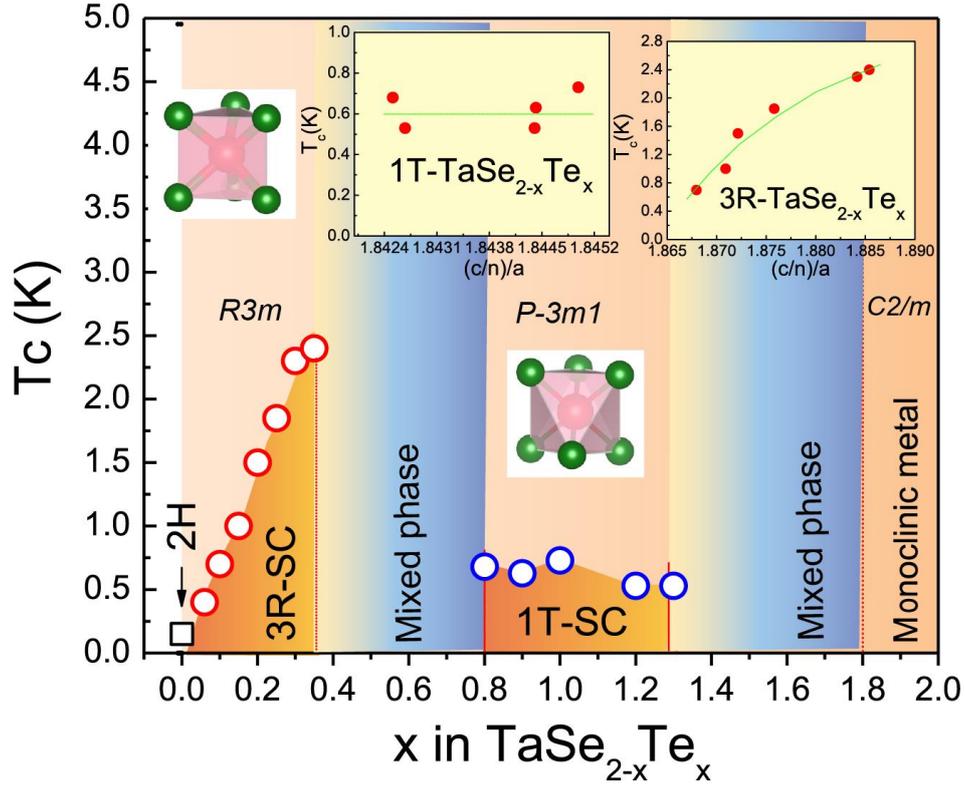

**Figure 5**. **Structural and Electronic phase diagram for TaSe$_{2-x}$Te$_x$.** Main panel – The composition stability ranges of the 2H, 3R, 1T and monoclinic MX$_2$ forms in TaSe$_{2-x}$Te$_x$. The TaX$_2$ coordination polyhedra are highlighted. Single phase regions are shown in pink, and multiple phase regions are shown in blue. The variation of the superconducting transition temperature with $x$ is also shown, as are the. Inset (left) The variation of the superconducting T$_c$ with the reduced $c/a$ ratio ($c/n$)/$a$, for 3R-TaSe$_{2-x}$Te$_x$ (0.1 ≤ x ≤ 0.35), and 1T-TaSe$_{2-x}$Te$_x$ (0.8 ≤ x ≤ 1.3). $n$ = number of layers in the stacking repeat, and $c$ and $a$ are the unit cell parameters.